\documentclass[journal=jpccck,manuscript=article]{achemso}
\usepackage{multirow}
\usepackage{chemformula} 
\usepackage[T1]{fontenc} 
\usepackage{hyperref}
\usepackage{comment}
\usepackage{amssymb}

\author{S. Mounbou}
\affiliation[Yaounde]
{Laboratory of Mechanics, Faculty of Science, University of Yaound\'e I, P.O. Box 812, Yaound\'e, Cameroon.}

\author{S. I. Fewo}
\affiliation[IFUnB]
{Laboratory of Mechanics, Faculty of Science, University of Yaound\'e I, P.O. Box 812, Yaound\'e, Cameroon.}

\author{L. A. Ribeiro, Jr}
\affiliation[IFUnB]
{University of Bras\'ilia, Institute of Physics, 70910-900, Bras\'ilia, Federal District, Brazil.}
\alsoaffiliation[LCCMat]
{Computational Materials Laboratory, LCCMat, Institute of Physics, University of Bras\'ilia, 70910-900, Bras\'ilia, Federal District, Brazil.}
\email{ribeirojr@unb.br}

\author{C. Kenfack-Sadem}
\affiliation[LAMACET]
{Laboratory of Condensed Matter-Electronics and Signal Processing (LAMACET), Department of Physics, Faculty of Science, University of Dschang, P.O. Box 67 Dschang, Cameroon.}

\title[Polaron-induced modifications]
  {Polaron-induced modifications in the linear and nonlinear optical properties of graphene under electric and magnetic fields}


\begin{document}

\begin{tocentry}
\begin{center}
\includegraphics[width=0.5\linewidth]{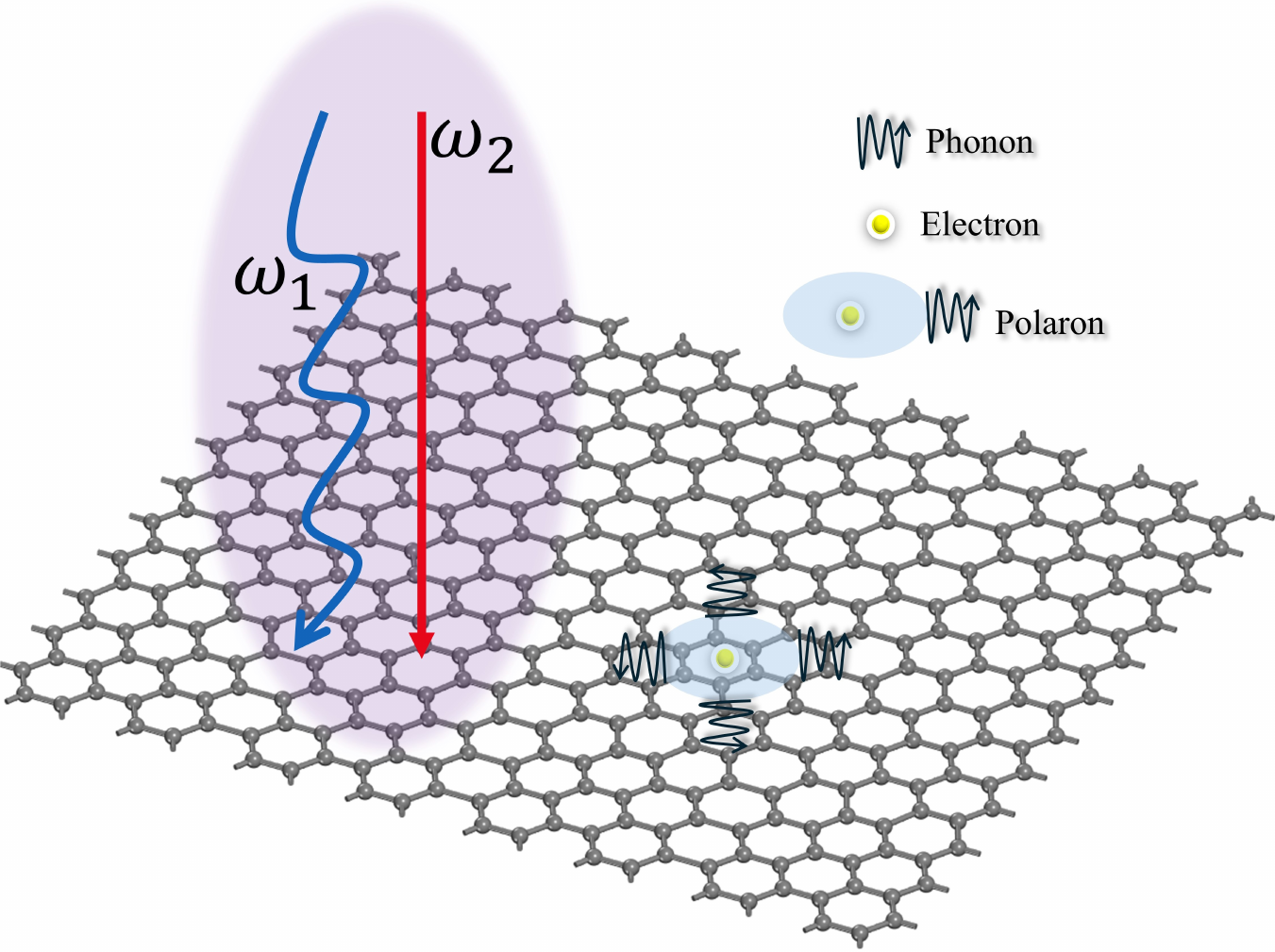}    
\end{center}
Polaron-induced modifications in the linear and nonlinear optical properties of graphene.
\end{tocentry}

\begin{abstract}
\noindent Polarons are the primary charge carriers in organic materials. A deep understanding of their properties can open channels for novel optoelectronic applications. By applying electric and magnetic fields, we investigate the influence of polaron interactions on the linear and nonlinear optical properties of a graphene monolayer between a substrate and air. Using the density matrix approach, we derive the linear and nonlinear optical absorption coefficients and the relative refractive index by incorporating the zero-energy level. Our numerical results reveal that the polaron effect and the magnetic field induce shifts in the peak positions of the optical absorption coefficients and refractive index. Moreover, while the presence of electric and magnetic fields significantly alters the amplitude of the absorption coefficients, only the magnetic field affects the refractive index amplitude. Additionally, we find that the magnetic field amplifies the influence of surface optical phonons on the optical properties of graphene. These findings provide deeper insights into the optical behavior of graphene in external fields, which could be relevant for optoelectronic applications.
\end{abstract}

\section{Introduction}
Graphene, a two-dimensional material, exhibits exceptional nonlinear optical properties due to its unique band structure and interband optical transitions across a wide range of photon energies \cite{liu2012polaron,feng2010graphene,mikhailov2007nonlinear}. A key feature of graphene in a magnetic field is its unconventional Landau levels \cite{nguyen2017optical}. The valence and conduction bands meet at the zero Landau level, allowing electrons and holes to coexist without an energy gap \cite{yang2010nonlinear,wang2015optical}. Various methods \cite{kandemir2018polaron,kandemir2017nonlinear,syzranov2008magnetopolaron,fistul2007polaron,song2013levitov,bokdam2014firstprinciples,hague2014polaron,rengel2014optical,scharf2013spin,konar2010charged} have been proposed to induce an energy gap at this level, including the interaction between graphene’s charge carriers and the vibration modes of a polar substrate \cite{wang2015optical}. These interactions modify the material’s linear dispersion, opening the gap due to the polaron effect \cite{wang2015optical}. Experimental studies using magneto-optical spectroscopy \cite{chen2014graphene} have confirmed the emergence of such an energy gap in graphene when placed on a substrate.

The linear and nonlinear optical properties of low-dimensional semiconductors, including graphene \cite{nguyen2017optical}, transition metal dichalcogenides \cite{nguyen2018field}, and double parabolic quantum wells \cite{ungan2019optical,ungan2014nonlinear}, have been extensively studied under the influence of electric and magnetic fields, considering interband, extraband, and mixed transitions. These studies reveal that external fields strongly influence the optical response of such materials. However, the zero Landau level of graphene has often been overlooked in these analyses due to the absence of an intrinsic energy gap. Recent research \cite{wang2015optical} suggests that a gap can emerge at this level due to the polaron effect, raising fundamental questions about possible optical transitions in this regime. Investigating these transitions necessitates an in-depth study of key optical properties, such as the optical absorption coefficient and relative refractive index. Charge carrier confinement techniques in low-dimensional systems have proven effective in enhancing these optical properties \cite{dakhlaoui2015linear}.

In graphene, charge carriers behave as Dirac fermions \cite{matulis2007weiss,demartino2007magnetic,peres2006dirac}, exhibiting chiral properties that lead to unconventional transport phenomena. One such effect is the symmetric transmission of charge carriers across a sufficiently high electrostatic potential, which prevents their confinement by electrostatic barriers \cite{biswas2010ballistic,masir2008direction,wu2008transport}. To address this limitation, several techniques have been proposed \cite{gonzalez2009resonant,biswas2010biased,bliokh2010tunable}, including the application of a perpendicular magnetic field \cite{matulis2007weiss,demartino2007magnetic}, magnetic barriers \cite{demartino2007magnetic,peres2006dirac,biswas2010ballistic,masir2008direction,wu2008transport}, and an in-plane electric field \cite{wu2008transport,gonzalez2009resonant}. Additionally, the interaction between charge carriers and the vibrational modes of a polar substrate has been identified as a mechanism for inducing confinement \cite{wang2015optical,li2010surface}. A perpendicular magnetic field modifies the linear dispersion of Dirac particles, with their energy scaling as the square root of the field intensity \cite{matulis2007weiss}. Meanwhile, an in-plane electric field modulated by magnetic barriers can induce highly asymmetric charge carrier transmission, enabling confinement even at normal incidence \cite{biswas2010ballistic}.
hu2015hybrid,
Studies on the polaron effect in quantum dots \cite{hu2015hybrid,wu2013polaron,parvathi2013polaron} under electric \cite{hu2015hybrid} and magnetic fields \cite{wu2013polaron} have demonstrated that charge carrier-phonon interactions enhance confinement, thereby facilitating optical transitions between energy levels. As a result, the peaks of linear and nonlinear optical absorption coefficients shift, and the refractive index is modified due to the polaronic effect, which is influenced by external fields \cite{hu2015hybrid,wu2013polaron}. These findings suggest that materials exhibiting strong photophysical properties \cite{mueller2010graphene} and significant optical nonlinearities hold potential for applications in high-speed optical communication and optical limiting technologies \cite{wang2009broadband,feng2010nonlinear}.

In this study, we theoretically investigate the influence of polaron interactions on the linear and nonlinear optical properties of a graphene monolayer placed on a substrate under the effect of electric and magnetic fields. Our analysis focuses on the optical absorption coefficient and relative refractive index. The magnetic field is assumed to be translationally invariant along the y-direction, while the electric field is applied within the graphene plane. We employ the Lee-Low-Pines formalism to determine the ground-state energy, followed by the compact density matrix approach to evaluate the optical properties. The manuscript is structured as follows: Section 2 describes the theoretical model and calculations, Section 3 presents the numerical results and discussion, and Section 4 concludes the study.

\section{Methodology}

\subsection{Model Hamiltonian}

Consider a graphene monolayer in the $(x,y)$ plane on which a substrate rests in the presence of 
an electric field applied \cite{biswas2010ballistic} in the $x$-direction and influenced by an inhomogeneous magnetic \cite{demartino2007magnetic} field normal to the plane of the graphene, invariant along the $y$-direction and varying along the $x$-axis, $\vec{B} = B(x,y) \hat{e}_z$. The vector potential is chosen such that $\vec{A}(x,y) = A(x) \hat{e}_y$. By considering the substrate vibrational modes and their interaction with the charge carriers of the graphene, the Hamiltonian is written as

\begin{equation}
\displaystyle H = H_e + V(r) + H_{ph} + H_{e-ph},
\label{hamiltonian}
\end{equation}

\noindent with

\begin{equation}
\displaystyle H_e = v_F \vec{\sigma} \cdot \left( \vec{p} + \frac{e}{c} \vec{A}(x,y) \right),
\end{equation}

\noindent where $v_F$ is the Fermi velocity, $\vec{\sigma}$ represents the Pauli matrices, and $\vec{p}$ is the momentum operator.

\begin{equation}
\displaystyle V(r) =
\begin{cases} 
    -eFx, & \text{for } -d \leq x \leq d \\
    0, & \text{otherwise}
\end{cases}
\end{equation}

\begin{equation}
\displaystyle H_{ph} = \sum_{k,\nu} \hbar \omega_{SO,\nu} a_k^{+} a_k,
\end{equation}

\noindent and

\begin{equation}
\displaystyle H_{e-ph} = \sum_{k,\nu} \left( M_{k,\nu} a_k e^{i \vec{k} \cdot \vec{r}} + M^*_{k,\nu} a_k^+ e^{-i \vec{k} \cdot \vec{r}} \right).
\end{equation}

The first term in Eq.~(\ref{hamiltonian}) describes the electron momentum energy, where $v_F$ is the
Fermi velocity, $\vec{\sigma} = (\sigma_x, \sigma_y)$ are the $2 \times 2$ Pauli matrices, and $c$ is the speed of light.
For a square well magnetic barrier~\cite{demartino2007magnetic}, where $\vec{B} = B_0 \hat{e}_z$ (with constant $B_0$) within the strip 
$-d \leq x \leq d$ but $B = 0$ otherwise,

\begin{equation}
\displaystyle B(x, y) = B_0 \theta(d^2 - x^2),
\end{equation} 

\noindent where $\theta$ is the Heaviside step function. The magnetic vector potential is written as

\begin{equation}
\displaystyle A(x) = \frac{c}{e \times l_B^2} x,
\end{equation} 

\noindent with $\ell_B = \left(\frac{\hbar}{e B_0} \right)^{1/2}$, which is the magnetic length.

The second term is the potential profile due to a homogeneous electric field $F$ applied 
between $x = -d$ and $x = d$ . The third term stands for the surface optical (SO) phonon with energy 
$\hbar \omega_{\nu} \left( \nu = SO_1, SO_2 \right)$ including SO mode, and $a_k^+ (a_k)$ is the annihilation (creation) operator for 
the phonon with wave vector $\vec{k}$. The fourth term $H_{e-ph}$ is the coupling between the carriers and 
SO phonon modes with the coupling matrix $M_{k,\nu}$. In addition, the optical phonons of the 
substrate included in our model have two longitudinal branches corresponding to the 
frequencies $\hbar \omega_{SO_1}$ and $\hbar \omega_{SO_2}$~\cite{wang2015optical}. The coupling matrix for the SO phonon induced by the polar 
substrate is~\cite{wang2015optical}

\begin{equation}
\displaystyle M_{k,\nu} = \sqrt{\frac{Q^2 \eta \hbar \omega_{SO,\nu}}{2 \varepsilon_0 k}} \exp(-kz).
\end{equation} 

\noindent $Q$ is the electric charge and 
\begin{equation}
\displaystyle \eta = \frac{\kappa_0 - \kappa_{\infty}}{(\kappa_0 + 1)(\kappa_{\infty} + 1)}
\end{equation}
\noindent is the ratio of known dielectric constants of substrates that indicates polarization resistance. $\varepsilon_0$ is the permittivity of the vacuum, and $\kappa_{\infty} (\kappa_0)$ is the high (low) frequency dielectric constant. $Z_0$ is the internal distance between the graphene monolayer and the polar substrate.

\subsection{Ground state energy}

The Hamiltonian of Eq.~(\ref{hamiltonian}) is transformed using the Lee-Low-Pines formalism~\cite{wang2015optical}

\begin{equation}
\displaystyle
\begin{cases}
S_1 = \exp \left[ -i \sum_k k r a_k^{+} a_k \right] \\
S_2 = \exp \left[ \sum_k \left( \phi_k a_k^{+} - \phi_k^* a_k \right) \right]
\end{cases}
\end{equation} 

\noindent as well as the position and momentum operators defined by~\cite{wang2015optical}

\begin{equation}
\displaystyle
\begin{cases}
r_j = \left( \frac{i}{\lambda \sqrt{2}} \right) (b_j - b_j^{+}) \\
p_j = \left( \frac{\hbar \lambda}{\sqrt{2}} \right) (b_j^{+} + b_j) 
\end{cases},
\end{equation} 

\noindent where $\lambda = \left( \frac{e B_0}{2 \hbar} \right)^{1/2}$, the index $j = x, y$. 
$\phi_k$ and $\phi_k^*$ are variational functions. $b_j$ are bosonic operators obeying the commutation relation. We obtain a new Hamiltonian by applying the relation $H' = S_2^{-1} S_1^{-1} H S_1 S_2$, which can be written as

\begin{equation}
\begin{aligned}
\displaystyle H' = v_F \sigma_x \left[ \frac{\hbar \lambda}{\sqrt{2}} (b_x^+ + b_x) + \frac{1}{\lambda \sqrt{2} \times l_B^2} (b_x - b_x^+) - \sum_k \hbar k_x (a_k^+ + \phi_k^*)(a_k + \phi_k) \right] \\
+ v_F \sigma_y \left[ \frac{\hbar \lambda}{\sqrt{2}} (b_y^+ + b_y) - \sum_k \hbar k_y (a_k^+ + \phi_k^*)(a_k + \phi_k) \right] + V(r) \\
+ \sum_{k,\nu} \hbar \omega_{SO,\nu} (a_k^+ + \phi_k^*)(a_k + \phi_k) + \sum_{k,\nu} M_{k,\nu} (a_k + \phi_k) \cdot \exp \left[ -\frac{k^2}{2\lambda^2} \right] \\
\cdot \exp \left[ \frac{k_j b_j^+}{\lambda \sqrt{2}} \right] \cdot \exp \left[ -\frac{k_j b_j}{\lambda \sqrt{2}} \right] \\
+ \sum_{k,\nu} M_{k,\nu}^* (a_k^+ + \phi_k^*) \cdot \exp \left[ -\frac{k^2}{2\lambda^2} \right] 
\cdot \exp \left[ -\frac{k_j b_j^+}{\lambda \sqrt{2}} \right] \cdot \exp \left[ \frac{k_j b_j}{\lambda \sqrt{2}} \right].
\end{aligned}
\end{equation}

The energy of $n$ states is obtained by applying the relation

\begin{equation}
\displaystyle E_n = \langle 0 | \langle \psi_n | H' | \psi_n \rangle | 0 \rangle,
\end{equation}

\noindent with

\begin{equation}
\displaystyle | \psi_n \rangle | 0 \rangle = \frac{1}{\sqrt{2}} 
\begin{pmatrix}
\displaystyle C_n | n-1 \rangle | 0 \rangle \\
\displaystyle C_n | n \rangle | 0 \rangle 
\end{pmatrix}.
\end{equation}

Here, $C_n = 1$ for $n = 0$ and $C_n = \frac{1}{\sqrt{2}}$ for $n \neq 0$. 
$|\psi_n\rangle$ is the electron wave function for the $n$ states $(n = 0,1...)$. 
$|0\rangle$ is the zero state of the phonon verifying the relations $a_k |0\rangle = 0$, 
$b_j |0\rangle = 0$, and $b_j^+ |0\rangle = 1$.

For the state $n = 0$ (corresponding to the ground state), we obtain

\begin{equation}
\begin{aligned}
\displaystyle E_0 = \pm \Bigg\{ & \sum_{k,\nu} \hbar k \phi_k^* \phi_k + E(r) 
+ \sum_{k,\nu} \hbar \omega_{SO,\nu} \phi_k^* \phi_k \\
& + \sum_{k,j} \Big( M_{k,\nu} \phi_k + M_{k,\nu}^* \phi_k^* a_k + \phi_k \Big) 
\cdot \exp \left( -\frac{k^2}{2\lambda^2} \right) \\
& \cdot \left( 1 + \frac{k^2}{4\lambda^2} \right) \Bigg\},
\end{aligned}
\end{equation}

\noindent with

\begin{equation}
\displaystyle \phi_k = \frac{-M_{k,\nu}^*}{4\lambda^2 \left( v_F \hbar k + \hbar \omega_{SO,\nu} \right)} 
\cdot \exp \left( -\frac{k^2}{2\lambda^2} \right) 
\cdot \left( k^2 + 4\lambda^2 \right).
\end{equation} 

\noindent $E_r = \left\langle -e F x \right\rangle$, the symbol $\langle ... \rangle$ denoting the average for the wave function $\psi_0(x)$ given by~\cite{nguyen2017optical}

\begin{equation}
\displaystyle  \psi_0(x) = \frac{1}{\sqrt{\pi^{1/2} \ell_B}} \exp \left( \frac{-1}{2\ell_B^2} x^2 \right).
\end{equation} 

\noindent Finally, the ground state energy is written as follows:

\begin{equation}
\begin{aligned}
\displaystyle E_0 = \pm \Bigg\{ & \sum_{\nu} \int dk \,
\frac{Q^2 \eta \hbar \omega_{SO,\nu} (k^2 + 4\lambda^2)}
{64\pi \varepsilon_0 \lambda^4 (v_F \hbar k + \hbar \omega_{SO,\nu})} \\
& \times \exp \left( -\frac{k^2}{\lambda^2} - 2k z_0 \right) 
+ \frac{e F \ell_B^2}{\sqrt{\pi}} (\exp(-4) -1) \Bigg\}.
\end{aligned}
\label{E0}
\end{equation} 

\subsection{Linear and nonlinear optical properties}

The linear and nonlinear optical properties, such as optical absorption coefficient and relative 
refractive indexes are calculated using the compact density matrix approach~\cite{keshavarz2010linear,guo2015linear,al2015effects}.

\subsubsection{Linear and nonlinear optical absorption coefficient}

Knowing the energy branches $E_{0+}$ and $E_{0-}$ due to the polaron effect given by Eq.~(\ref{E0}) and using the dipolar electric matrix element $M_{\alpha\alpha'} = e \langle \psi_{\alpha'} | x | \psi_{\alpha} \rangle$ ($\alpha' = 0^+$ and $\alpha = 0^-$), the linear and nonlinear optical absorption coefficients are given, respectively, by

\begin{equation}
\displaystyle  \chi^{(1)} (\Omega) = \Omega \sqrt{\frac{\mu}{\varepsilon_r}} 
\frac{|M_{\alpha\alpha'}|^2 n_e \hbar \Gamma_0}
{(\Delta E - \hbar \Omega)^2 + (\hbar \Gamma_0)^2}.
\label{chi1}
\end{equation} 

\noindent and

\begin{equation}
\displaystyle \chi^{(3)}(\Omega, I) = -2\Omega \sqrt{\frac{\mu}{\varepsilon_r}} 
\left( \frac{I}{\varepsilon_0 n_r c} \right)
\frac{|M_{\alpha\alpha'}|^4 n_e \hbar \Gamma_0}
{\left[(\Delta E - \hbar \Omega)^2 + (\hbar \Gamma_0)^2\right]^2},
\label{chi2}
\end{equation} 

\noindent where $\hbar \Omega$ is the photon energy, $\mu$, is the magnetic permeability, $\varepsilon_r$ is the relative permittivity, $n_e$ is the charge density, $\Gamma_0$ is the phenomenological relaxation rate, $\Delta E = E_{0+} - E_{0-}$ is the energy difference of the two bands, $I$ is the optical intensity of the photon which causes the optical transition between the two bands, $c$ is the light velocity, $\varepsilon_0$ is the vacuum permittivity, and $n_r$ is the refractive index. Starting from Eqs.~(\ref{chi1}) and (\ref{chi2}), we obtain the total optical absorption coefficient given by

\begin{equation}
\displaystyle \chi(\Omega, I) = \chi^{(1)}(\Omega) + \chi^{(3)}(\Omega, I).
\end{equation} 

\subsubsection{Linear and nonlinear refractive index}

Using the same procedure as in the case of the optical absorption coefficient, we obtain the linear and third-order nonlinear refractive index given, respectively, by

\begin{equation}
\displaystyle \frac{\Delta n^{(1)}(\Omega)}{n_r} =
\frac{|M_{\alpha\alpha'}|^2 n_e}{2n_r^2 \varepsilon_0} 
\left[ \frac{\Delta E - \hbar \Omega}{(\Delta E - \hbar \Omega)^2 + (\hbar \Gamma_0)^2} \right].
\end{equation} 

\noindent and

\begin{equation}
\displaystyle \frac{\Delta n^{(3)}(\Omega, I)}{n_r} =
\frac{- \mu c |M_{\alpha\alpha'}|^4 n_e I (\Delta E - \hbar \Omega)}
{n_r^3 \varepsilon_0 \left[ (\Delta E - \hbar \Omega)^2 + (\hbar \Gamma_0)^2 \right]^2}.
\end{equation} 

\noindent The following relation then gives the total refractive index:

\begin{equation}
\displaystyle \frac{\Delta n(\Omega, I)}{n_r} =
\frac{\Delta n^{(1)}(\Omega)}{n_r} + \frac{\Delta n^{(3)}(\Omega, I)}{n_r}.
\end{equation} 

The optical properties are given by Eqs.~(16) to (21) and depend strongly on the parameters of the graphene, the incident photon energy $\hbar \Omega$, and the energy gap $\Delta E$ due to the polaron effect. These equations do not clearly show the polaron's effect on its parameters. Hence, numerical analyses are necessary.

\section{Results}

In this section, we analyze the influence of the polaron effect on the linear and nonlinear optical properties of a graphene monolayer as a function of incident photon energy. The parameters of the substrates used in the simulations are listed in Table~\ref{tab:phonon}, while additional simulation parameters are adopted from previous studies~\cite{nguyen2017optical}: $n_e = 10^{13} \, cm^{-2}$, $n_r = 2.0$, $I = 5 \, MW / m^2$, $\hbar \Gamma_0 = 0.2 \sqrt{B} \, meV$, $Q = 100$, $k_c = 0.5 \, m^{-1}$, and $z_0 = 1 \, nm$.

\begin{table}[!htb]
\centering
\caption{Surface-optical phonon modes parameters for different polar substrates~\cite{wang2015optical}.}
\label{tab:phonon}
\begin{tabular}{|c|c|c|c|c|}
\hline
\textbf{Quantity (units)} & $h-BN$ & $SiC$ & $SiO_2$ & $HfO_2$ \\ \hline
$\hbar \omega_{SO,\nu} \, (meV)$ & 296 & 283 & 206 & 72 \\ \hline
$\eta$ & 0.032 & 0.04 & 0.08 & 0.12 \\ \hline
\end{tabular}
\end{table}

\begin{figure}[!htb]
    \centering
    \includegraphics[width=\linewidth]{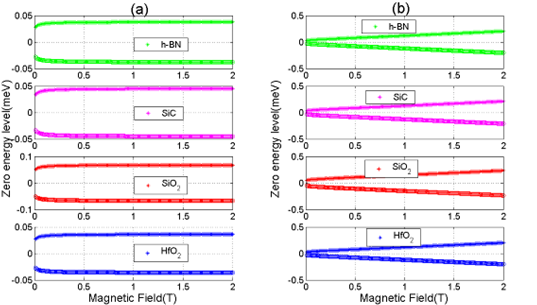}
    \caption{Zero-energy Landau level branches ($E_{0+}$ and $E_{0-}$) as a function of magnetic field strength for different polar substrates: \textit{h}-BN, SiC, SiO$_2$, and HfO$_2$. (a) Without and (b) with an applied in-plane electric field ($F = 0.3$~V/nm).}
    \label{fig1}
\end{figure}

Figure~\ref{fig1} shows the evolution of the zero-energy Landau level branches ($E_{0+}$ and $E_{0-}$) as a function of magnetic field strength for four polar substrates: $h$-BN, SiC, $SiO_2$, and $HfO_2$. Panels (a) and (b) correspond to the cases without ($F = 0$) and with ($F = 0.3$~V/nm) an applied in-plane electric field, respectively.

In the absence of the electric field (Figure~\ref{fig1}(a)), the energy splitting between the branches increases sharply for low magnetic field values and then saturates for $B \gtrsim 1$~T. This behavior is a direct signature of the polaronic gap opening at the zero Landau level, driven by interactions between graphene charge carriers and the substrate's surface optical (SO) phonons. The magnitude of this gap depends on two key substrate parameters: the phonon energy $\hbar\omega_{SO}$ and the polarization parameter $\eta$, which quantifies the strength of the electron-phonon coupling.

The substrates do not follow a simple trend. For example, although $HfO_2$ has the largest $\eta$ among the materials considered (suggesting strong coupling), it exhibits the smallest energy gap. In contrast, $SiO_2$ shows the widest splitting despite a lower $\eta$. This apparent contradiction highlights a subtle competition: a higher $\eta$ enhances coupling, but a low $\hbar\omega_{SO}$ reduces the phonon-mediated energy shift. In the case of $HfO_2$, its low value $\hbar\omega_{SO} = 72$meV (compared to 206meV for $SiO_2$) weakens the polaronic effect despite stronger polarization. Thus, both parameters must be considered jointly to predict the resulting energy shifts.

With the electric field applied (Figure~\ref{fig1}(b)), the splitting becomes more pronounced for all substrates. The electric field breaks the in-plane symmetry and enhances the localization of the carriers, reinforcing the interaction with SO phonons. We observe that the electric field refines the energy for low values of the magnetic field. For large values of the magnetic field the energy is constant for $F = 0$ but increases for $F$ other than 0. This cooperative effect between electric and magnetic fields amplifies the polaron-induced energy gap, providing a tunable mechanism for gap engineering.

These observations are consistent with prior studies on polaron effects in low-dimensional systems such as quantum dots~\cite{hu2015hybrid,wu2013polaron}, where similar external field configurations lead to enhanced optical transitions. Importantly, the formation of the gap facilitates interband transitions at the zero Landau level, implying that surface polar phonons can activate optical transitions in otherwise symmetry-protected regimes. This makes them key enablers for modulating the linear and nonlinear optical response of graphene-based devices.

\begin{figure}[!htb]
    \centering
    \includegraphics[width=\linewidth]{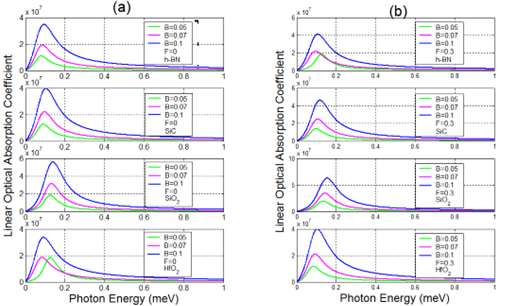}
    \caption{Linear optical absorption coefficient as a function of photon energy for different polar substrates: \textit{h}-BN, SiC, SiO$_2$, and HfO$_2$, under various magnetic field strengths ($B = 0.05$, $0.07$, and $0.1$~T). (a) Absorption spectra without electric field ($F = 0$); (b) with in-plane electric field ($F = 0.3$~V/nm).}
    \label{fig2}
\end{figure}

Figure~\ref{fig2} presents the variations in the linear optical absorption coefficient as a function of photon energy for four different polar substrates: $h$-BN, SiC, $SiO_2$, and $HfO_2$. Panel (a) shows the absorption response in the presence of the polaron effect alone ($F = 0$), while panel (b) includes the influence of an applied in-plane electric field ($F = 0.3$~V/nm). In each case, results are shown for three different magnetic field strengths: $B = 0.05$, $0.07$, and $0.1$~T.

In Figure~\ref{fig2}(a), which represents the case without an applied electric field ($F = 0$), the linear optical absorption spectra reveals the influence of the magnetic field on the polaron-induced optical transitions. For all substrates, increasing the magnetic field from 0.05T to 0.1T results in a gradual blue shift of the absorption peaks and a noticeable increase in peak intensity. This behavior is consistent with widening the energy gap at the zero Landau level due to the enhanced magnetic confinement of the charge carriers. As the magnetic field increases, more photon energy is required to excite interband transitions, resulting in the observed shift. At the same time, the stronger localization leads to more significant overlap between electron and phonon wavefunctions, thereby enhancing the oscillator strength of the transitions. Among the four substrates, $SiO_2$ and SiC exhibit the most pronounced absorption enhancement, likely due to their balanced $\eta$ and $\hbar\omega_{\mathrm{SO}}$, which optimizes the electron-phonon coupling strength. In contrast, $HfO_2$ shows the weakest peak response, further highlighting the importance of substrate phonon energy in determining the magnitude of the polaron effect, even when the polarization parameter is high.

On the other hand, in Figure~\ref{fig2}(b), where an in-plane electric field of $F = 0.3$~V/nm is applied, the absorption spectra exhibit an even more pronounced response to the magnetic field. Compared to the zero-field case, the absorption peaks become sharper, stronger, and more blue-shifted, confirming the cooperative role of electric and magnetic fields in modulating the polaronic dynamics. The electric field enhances carrier localization in the graphene sheet, which increases their interaction with surface optical phonons and further strengthens the polaron binding. As a result, the polaron-induced energy gap widens more rapidly with magnetic field strength, requiring higher photon energies to activate optical transitions. This leads to more intense absorption and broader spectral features. Again, $SiO_2$ and SiC stand out with the highest absorption coefficients, confirming their optimal combination of polarization strength and phonon energy. The case of $HfO_2$ remains an exception, with comparatively weaker absorption even under the combined influence of $B$ and $F$, due to its low $\hbar\omega_{\mathrm{SO}}$, which limits the available phonon-mediated transition energy despite its high $\eta$ value. These results suggest that external fields can be effectively used to tune the optical response of graphene–substrate heterostructures, especially when the substrate supports strong and energetically favorable polaronic interactions.

In all scenarios, the absorption coefficient increases with photon energy, reaching a well-defined peak before gradually decreasing. This bell-shaped response is characteristic of interband optical transitions near the zero Landau level, and both the magnetic and electric fields strongly influence the peak intensity and position.

Two key effects are observed as the magnetic field increases: (i) the peak intensity rises, and (ii) the peak position shifts toward higher photon energies. This trend reflects the widening of the polaron-induced energy gap under stronger magnetic confinement, which increases the energy required for optical transitions. These findings align with prior studies on magneto-polarons in confined systems~\cite{hu2015hybrid,wu2013polaron}, where similar field-induced shifts and broadenings were reported.

When the electric field is applied (Figure~\ref{fig2}(b)), the absorption peaks become broader and more intense across all substrates. This effect results from the enhanced localization of charge carriers and the strengthening of their coupling with substrate SO phonons under combined electric and magnetic fields. In particular, the field-enhanced interaction increases polaron binding energy, which contributes to the greater optical transition probability observed.

The relative behavior among the substrates is also notable. Substrates such as $SiO_2$ and $SiC$, which exhibit stronger effective polaronic coupling (balanced $\eta$ and $\hbar\omega_{\mathrm{SO}}$), show the most pronounced absorption enhancement. In contrast, $HfO_2$ demonstrates a weaker response, likely due to its low phonon energy reducing the effectiveness of the polaron formation despite its high polarization parameter.

\begin{figure}[!htb]
    \centering
    \includegraphics[width=\linewidth]{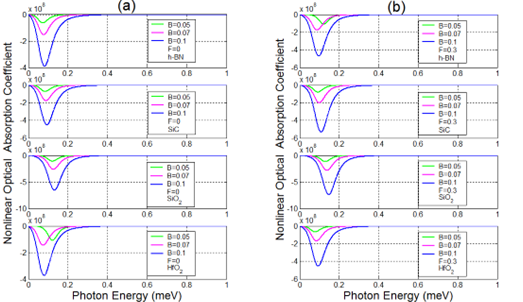}
    \caption{Third-order nonlinear optical absorption coefficient as a function of photon energy for different polar substrates: \textit{h}-BN, SiC, SiO$_2$, and HfO$_2$. (a) Without electric field ($F = 0$); (b) with in-plane electric field ($F = 0.3$~V/nm).}
    \label{fig3}
\end{figure}

Figure~\ref{fig3} displays the third-order nonlinear optical absorption coefficient as a function of photon energy for four different polar substrates: \textit{h}-BN, SiC, $SiO_2$, and $HfO_2$. Panel (a) corresponds to the case without an electric field ($F = 0$), while panel (b) includes the effect of an applied in-plane electric field ($F = 0.3$~V/nm). Both panels show results for increasing magnetic field strengths ($B = 0.05$, $0.07$, and $0.1$~T).

Across all panels, the absorption coefficient exhibits a negative peak (typical for saturable absorption) that becomes sharper and more intense as the magnetic field increases. This behavior originates from the enhancement of the polaron-induced energy gap, which modifies the material's nonlinear susceptibility. As photon energy increases, the absorption coefficient decreases from zero, reaches a minimum, and returns toward zero, revealing a resonant-like nonlinear response around the polaron transition energy.

In Figure~\ref{fig3}(a), without the electric field, this nonlinear signature becomes progressively more pronounced with increasing magnetic field. The larger the field, the stronger the confinement and, thus, the stronger the polaronic response. This trend increases the population difference between Landau levels and enhances the system’s third-order susceptibility. As observed in linear absorption (Fig.~\ref{fig2}), $SiO_2$ and SiC substrates produce the deepest and broadest nonlinear absorption minima, indicating stronger light-matter interaction due to their balanced $\eta$ and $\hbar\omega_{\mathrm{SO}}$ values. In contrast, $HfO_2$ displays the weakest response because of its low phonon energy despite a high $\eta$.

When the electric field is introduced (Figure~\ref{fig3}(b)), the nonlinear absorption coefficient becomes more negative for all substrates. This behavior confirms that the electric field enhances carrier localization and polaron binding, thereby increasing the effective nonlinear optical response. The shifts in the absorption minima toward higher photon energies mirror the broadening of the energy gap seen in the linear regime, demonstrating that external fields can simultaneously modulate both linear and nonlinear optical characteristics.

These findings are consistent with earlier studies on nonlinear optical behavior in polaronic systems under external fields~\cite{hu2015hybrid,wu2013polaron} and underscore the potential of field-tunable polar substrates to enhance the nonlinear optoelectronic response of graphene-based heterostructures.

\begin{figure}[!htb]
    \centering
    \includegraphics[width=\linewidth]{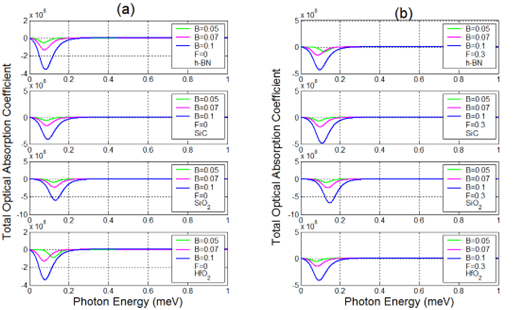}
    \caption{Total optical absorption coefficient (sum of linear and third-order nonlinear components) as a function of photon energy for different polar substrates: \textit{h}-BN, SiC, SiO$_2$, and HfO$_2$. (a) Without electric field ($F = 0$); (b) with in-plane electric field ($F = 0.3$~V/nm).}
    \label{fig4}
\end{figure}

Figure~\ref{fig4} illustrates the total optical absorption coefficient, which combines both linear and third-order nonlinear contributions, as a function of photon energy for the same four polar substrates: \textit{h}-BN, SiC, $SiO_2$, and $HfO_2$. Panels (a) and (b) show the cases without ($F = 0$) and with ($F = 0.3$~V/nm) an applied in-plane electric field, respectively, for magnetic fields $B = 0.05$, $0.07$, and $0.1$~T.

The overall spectral shape is governed by the interplay between linear enhancement and nonlinear absorption suppression. For low photon energies, the nonlinear component dominates --- as previously observed in Figure~\ref{fig3} --- leading to a negative total absorption coefficient near the resonance. This negative region, often associated with saturable absorption or optical bleaching, indicates that the material becomes more transparent under intense illumination, a hallmark of strong nonlinear response.

In Figure~\ref{fig4}(a), which corresponds to the case without an applied electric field ($F = 0$), the total optical absorption coefficient exhibits a distinctive negative peak across all substrates and magnetic field values. This behavior results from the dominance of the third-order nonlinear contribution, which outweighs the linear absorption near the polaron resonance energy. As the magnetic field increases from 0.05T, the depth of the absorption dip becomes more pronounced and shifts toward higher photon energies, reflecting the progressive widening of the polaron-induced energy gap. The strongest negative response for the $SiO_2$ and SiC substrates is consistent with their favorable balance between the polarization parameter $\eta$ and surface phonon energy $\hbar\omega_{\mathrm{SO}}$. These substrates support stronger electron-phonon coupling, which enhances the nonlinear interaction. In contrast, $HfO_2$ again shows the weakest absorption variation due to its low phonon energy, which limits the efficiency of polaron formation. Overall, this panel reveals that even without an external electric field, the magnetic field alone is sufficient to activate a strong nonlinear regime in the system, with a substantial impact on the total optical absorption.

When an in-plane electric field of $F = 0.3$~V/nm is applied (see Figure~\ref{fig4}(b)), the total optical absorption spectra reveal a marked enhancement in both the depth and breadth of the negative absorption region across all substrates. This amplification reflects the synergistic action of electric and magnetic fields, which jointly increase charge carrier localization and strengthen the electron-phonon interaction. As a result, the nonlinear absorption term becomes even more dominant, and the absorption minima shift further toward higher photon energies. $SiO_2$ and SiC again exhibit the most intense response, confirming their superior ability to sustain field-enhanced polaron effects. In this configuration, the absorption dips remain negative over a broader spectral range, underscoring the field-tunable nature of the optical properties in graphene–substrate heterostructures. These findings suggest that applying an external electric field is a powerful mechanism to modulate the overall absorption profile, which is essential for designing active optoelectronic components where dynamic control of transmission and absorption is required.

As the magnetic field increases, the absorption minima become deeper and shift toward higher photon energies, consistent with the widening of the polaron-induced energy gap. This shift is observed in all substrates but is most pronounced in $SiO_2$ and SiC, where the combination of moderate $\eta$ and high $\hbar\omega_{\mathrm{SO}}$ yields the most substantial polaronic effects. In contrast, $HfO_2$ again exhibits a relatively modest response due to its lower phonon energy.

With the application of the electric field (Figure~\ref{fig4}(b)), the amplitude of the absorption dip increases, especially for higher $B$ values. This enhancement confirms the synergistic effect of electric and magnetic fields in amplifying the polaron binding and, consequently, the nonlinear optical response. In this combined field regime, the total absorption remains negative over a broader energy range, reinforcing the nonlinear term's dominance in the system's optical response.

\begin{figure}[!htb]
    \centering
    \includegraphics[width=\linewidth]{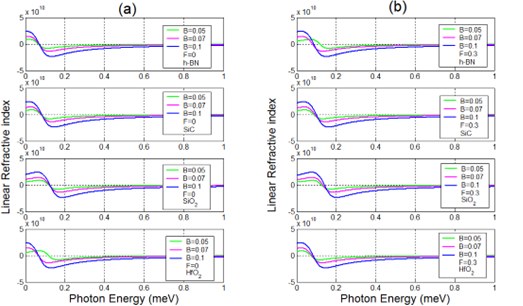}
    \caption{Linear relative refractive index as a function of photon energy for different polar substrates: \textit{h}-BN, SiC, SiO$_2$, and HfO$_2$. (a) Without electric field ($F = 0$); (b) with in-plane electric field ($F = 0.3$~V/nm).}
    \label{fig5}
\end{figure}

Figure~\ref{fig5} presents the variation of the linear relative refractive index as a function of photon energy for four different polar substrates: \textit{h}-BN, SiC, $SiO_2$, and $HfO_2$, under varying magnetic field strengths ($B = 0.05$, $0.07$, and $0.1$~T). Panel (a) shows the case without an applied electric field ($F = 0$), while panel (b) includes a field of $F = 0.3$~V/nm.

Without an electric field, the linear relative refractive index exhibits distinct dispersive features as shown in Figure~\ref{fig5}(a), with positive and negative peaks that become more pronounced as the magnetic field increases from 0.05T to 0.1T. These features are signatures of resonant optical behavior near the polaron-induced transition energy and reflect the underlying modification of the real part of the optical susceptibility due to the formation of magneto-polarons. The increase in the amplitude of these oscillations with stronger magnetic fields indicates enhanced polaron confinement, which modifies the group velocity of the transmitted light. Moreover, a substrate-dependent shift in the position of the resonance peaks is observed. Substrates such as $SiO_2$ and SiC, which possess favorable combinations of $\eta$ and $\hbar\omega_{\mathrm{SO}}$, display the most significant peak shifts, indicating a stronger interaction between the graphene carriers and substrate phonons. In contrast, $HfO_2$ shows less pronounced spectral displacement, consistent with its low phonon energy, which limits the influence of the polaron effect on the refractive response.

When the in-plane electric field is applied ($F = 0.3$\ref{fig5}(b), the overall structure of the linear relative refractive index remains qualitatively similar to the zero-field case. However, the resonance peaks undergo noticeable spectral shifts, especially for substrates with strong electron-phonon coupling. This behavior confirms that while the electric field has limited influence on the amplitude of the refractive index modulation, it plays a key role in altering the energy position of the resonant response. This shift arises from the field-induced enhancement of the polaron binding energy, which modifies the dispersion relation of charge carriers in the vicinity of the zero Landau level. The effect is particularly evident in $SiO_2$ and SiC, whose balanced $\eta$ and $\hbar\omega_{\mathrm{SO}}$ values allow for stronger substrate–carrier interaction. On the other hand, $h$-BN and $HfO_2$ exhibit smaller shifts, consistent with their weaker phonon-mediated coupling or lower phonon energies. These results reinforce the idea that electrical control of dispersion properties in graphene–substrate systems can be achieved by tuning both substrate choice and external field parameters.

In both panels, the curves exhibit a pronounced dispersive structure, with positive and negative extrema centered around specific photon energy values. These extrema correspond to resonant modifications in the optical response induced by transitions near the polaron gap. As the magnetic field increases, the magnitude of these features becomes more prominent, indicating that the refractive index is sensitive to magnetic confinement and the associated enhancement of the polaronic interaction.

Interestingly, the application of the electric field [Fig.~\ref{fig5}(b)] does not significantly alter the amplitude of the refractive index variations. This trend suggests that the real part of the susceptibility, which governs refraction, is less sensitive to electric-field-induced carrier localization than the absorption-related imaginary part. However, a clear shift in the peak positions is observed across all substrates, especially for those with higher polarization parameters (such as $SiO_2$ and SiC). This feature indicates that substrate phonons play a key role in modulating the dispersion characteristics and that the field-enhanced electron-phonon coupling modifies the resonance condition.

Among the four substrates, $SiO_2$ exhibits the largest shift and most pronounced dispersive behavior, consistent with its balanced $\eta$ and $\hbar\omega_{\mathrm{SO}}$. On the other hand, $HfO_2$, despite its high polarization parameter, shows a relatively less dramatic shift, once again highlighting the dampening role of low phonon energy.

\begin{figure}[!htb]
    \centering
    \includegraphics[width=\linewidth]{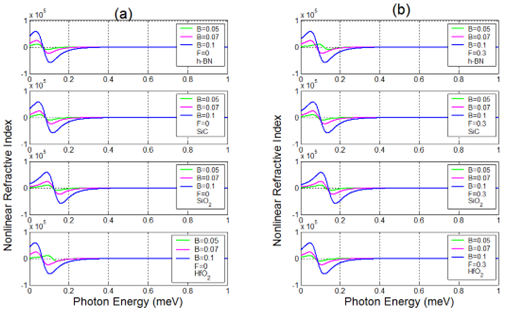}
    \caption{Third-order nonlinear relative refractive index as a function of photon energy for different polar substrates: \textit{h}-BN, SiC, SiO$_2$, and HfO$_2$. (a) Without electric field ($F = 0$); (b) with in-plane electric field ($F = 0.3$~V/nm).}
    \label{fig6}
\end{figure}

Figure~\ref{fig6} shows the behavior of the third-order nonlinear relative refractive index as a function of photon energy for four different polar substrates: \textit{h}-BN, SiC, $SiO_2$, and $HfO_2$, under varying magnetic field strengths ($B = 0.05$, $0.07$, and $0.1$~T). Panel (a) presents the case without an electric field ($F = 0$), and panel (b) includes an in-plane electric field of $F = 0.3$~V/nm.

In the absence of an electric field, the third-order nonlinear relative refractive index --- shown in Figure~\ref{fig6}(a) --- exhibits a clear dispersive pattern characterized by alternating positive and negative regions centered around the polaron resonance energy. As the magnetic field increases from 0.05T, the nonlinear response's amplitude and spectral width become more pronounced. This behavior reflects the enhanced formation of magneto-polarons and their influence on the nonlinear optical susceptibility of the system. Compared to the linear case (Fig.~\ref{fig5}(a)), the resonance features are more sensitive to changes in the magnetic field, resulting in greater peak shifts across all substrates. Among them, $SiO_2$ and SiC display the strongest nonlinear dispersive response, which is consistent with their efficient electron-phonon coupling due to favorable combinations of polarization parameter $\eta$ and phonon energy $\hbar\omega_{\mathrm{SO}}$. In contrast, $HfO_2$ shows a weaker and narrower nonlinear profile, again highlighting the limiting role of low phonon energy in shaping the polaronic effects. These results underscore that the nonlinear refractive index offers enhanced sensitivity to substrate-induced polaron dynamics even without an electric field, especially under moderate magnetic fields.

With the application of an in-plane electric field ($F = 0.3$\ref{fig6}(b) maintains a similar qualitative shape to those in the zero-field case but now exhibits more pronounced spectral shifts, especially for substrates with stronger electron-phonon coupling. The electric field enhances carrier localization in the graphene layer, reinforcing the interaction with substrate surface phonons and altering the effective polaronic energy levels. As a result, the resonance peaks shift to higher photon energies for all substrates, reflecting the field-induced modulation of the nonlinear dispersive response. The amplitude of the nonlinear refractive index does not increase substantially with the electric field, suggesting that its influence is primarily dispersive rather than absorptive in this context. Once again, $SiO_2$ and SiC demonstrate the largest field-induced shifts, while $HfO_2$ shows minimal response due to its low $\hbar\omega_{\mathrm{SO}}$. These observations highlight that although the nonlinear refractive index remains smaller in magnitude than its linear counterpart, it is more responsive to external field tuning, making it a valuable quantity for characterizing and controlling field-sensitive optical behavior in graphene–substrate systems.

Overall, the nonlinear refractive index displays a rich dispersive structure with alternating positive and negative regions, reflecting the resonant enhancement and suppression of the third-order optical susceptibility near the polaron transition energy. Compared to the linear case (Figure~\ref{fig5}), the shifts in the resonance features are more pronounced, especially under stronger magnetic fields. This increased sensitivity highlights the nonlinear refractive index as a more effective probe of the underlying polaronic dynamics and substrate interactions.

\begin{figure}[!htb]
    \centering
    \includegraphics[width=\linewidth]{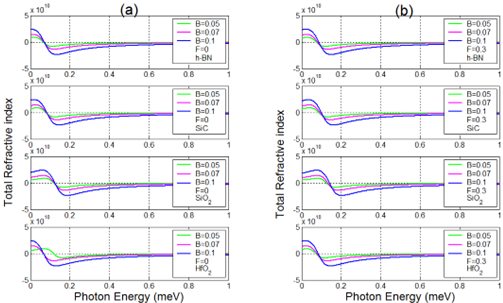}
    \caption{Total relative refractive index (sum of linear and third-order nonlinear contributions) as a function of photon energy for polar substrates: \textit{h}-BN, SiC, SiO$_2$, and HfO$_2$. (a) Without electric field ($F = 0$); (b) with in-plane electric field ($F = 0.3$~V/nm).}
    \label{fig7}
\end{figure}

Finally, Figure~\ref{fig7} displays the total relative refractive index as a function of photon energy for four different polar substrates: \textit{h}-BN, SiC, $SiO_2$, and $HfO_2$. This total response is obtained by summing the linear and third-order nonlinear refractive indices. Panel (a) shows the case without an electric field ($F = 0$), while panel (b) includes the effect of an in-plane electric field ($F = 0.3$~V/nm), for magnetic field values of $B = 0.05$, $0.07$, and $0.1$~T.

The total refractive index curves retain the general dispersive profile observed in the linear case, with both positive and negative shifts appearing near the polaron resonance energy. This trend is expected, as the linear contribution dominates the overall magnitude, while the nonlinear component acts as a perturbation that subtly shifts or distorts the features rather than overrides them. Nevertheless, the influence of the nonlinear term is still visible, particularly in the slight asymmetries and spectral displacements observed as the magnetic field increases.

Increasing the magnetic field in both panels results in a progressive shift of the refractive features toward higher photon energies, consistent with the broadening of the polaron-induced energy gap. Among the substrates, $SiO_2$ and SiC again exhibit the strongest modulation, reflecting their more efficient coupling between graphene charge carriers and surface optical phonons. Conversely, $HfO_2$ continues to show a weaker spectral response due to its low phonon energy despite its higher polarization parameter.

Without an applied electric field, the total relative refractive index—shown in Figure~\ref{fig7}(a)—reveals a dispersive structure that closely resembles the linear refractive index profile. This similarity confirms that the linear contribution dominates the overall refractive behavior of the system, while the nonlinear component introduces only minor modifications. As the magnetic field increases, the dispersive features become more pronounced and shift toward higher photon energies, consistent with the enhancement of the polaron-induced energy gap. These shifts are particularly visible for $SiO_2$ and SiC substrates, whose balanced values of $\eta$ and $\hbar\omega_{\mathrm{SO}}$ favor stronger electron-phonon coupling and thus more noticeable refractive changes. The substrates $h$-BN and $HfO_2$ exhibit weaker responses due to low $\eta$ or low phonon energy, respectively. While the nonlinear refractive index remains small in magnitude, it subtly distorts the total profile, slightly altering the symmetry and slope of the spectral features. This behavior indicates that even when small, nonlinear effects can influence light propagation characteristics in graphene-based systems under moderate magnetic fields.

With the electric field applied ($F = 0.3$\ref{fig7}(b), it continues to be primarily dictated by the linear component, but subtle field-induced modifications become evident. As in the zero-field case, the overall spectral shape retains the dispersive character, with resonant features shifting toward higher photon energies as the magnetic field increases. However, compared to panel (a), these shifts are slightly enhanced in substrates exhibiting stronger polaronic coupling, such as $SiO_2$ and SiC. This trend indicates that while the nonlinear contribution remains small, it becomes more influential under combined magnetic and electric fields, subtly altering the refractive landscape. In contrast, $h$-BN and $HfO_2$ show minimal change, reinforcing that significant modulation of the total refractive index requires strong substrate, carrier interaction, and an effective external driving mechanism. Panel (b) demonstrates that the electric field acts as a secondary control parameter, enabling fine-tuning refractive properties without drastically altering the dominant linear behavior, an asset for precision control in integrated photonic applications.

\section{Conclusions}

In summary, we have investigated the influence of the polaron effect on the linear and nonlinear optical properties of a graphene monolayer placed on various polar substrates and subjected to perpendicular magnetic and in-plane electric fields. We derived the absorption coefficient and the relative refractive index under realistic field conditions by employing the Lee–Low–Pines formalism to compute the ground-state energy and the density matrix approach to derive the optical responses.

Our results demonstrate that surface optical phonons from the substrate enable optical transitions at the zero Landau level of graphene. This interaction significantly enhances both the linear and third-order nonlinear absorption coefficients and the associated refractive index changes. Moreover, the combination of magnetic and electric fields intensifies the localization and confinement of charge carriers, strengthening polaron formation and leading to sharper, tunable optical responses—particularly in substrates with favorable dielectric and phonon properties.

These findings highlight the potential of field-assisted polaron engineering as a versatile tool to modulate the optical and transport properties of two-dimensional systems. As a next step, we intend to explore the role of acoustic polarons and their impact on carrier mobility and conductivity in graphene, extending the analysis beyond the optical regime and toward realistic applications in nanoelectronics and optoelectronics.

\begin{acknowledgement}
L.A.R.J. acknowledges the financial support from FAP-DF grants $00193.00001808$ $/2022-71$ and $00193-00001857/2023-95$, FAPDF-PRONEM grant $00193.00001247/2021-20$, PDPG-FAPDF-CAPES Centro-Oeste $00193-00000867/2024-94$, and CNPq grants $350176/2022-1$ and $167745/2023-9$. The authors gratefully acknowledge CAPES for the financial support provided for the publication of this work under the CC BY Open Access license through the ACS-CAPES agreement.
\end{acknowledgement}

\bibliography{references}

\end{document}